# Canonical and DLPNO-Based G4(MP2)XK-Inspired Composite Wave Function Methods Parametrized against Large and Chemically Diverse Training Sets: Are They More Accurate and/or Robust than Double-Hybrid DFT?

Emmanouil Semidalas and Jan M. L. Martin*

Read Online

ACCESS | Metrics & More | Article Recommendations | Supporting Information

**ABSTRACT:** The large and chemically diverse GMTKN55 benchmark was used as a training set for parametrizing composite wave function thermochemistry protocols akin to G4(MP2)XK theory (Chan, B.; Karton, A.; Raghavachari, K. *J. Chem. Theory Comput.* **2019**, *15*, 4478−4484). On account of their availability for elements H through Rn, Karlsruhe def2 basis sets were employed. Even after reparametrization, the GMTKN55 WTMAD2 (weighted mean absolute deviation, type 2) for G4(MP2)-XK is actually inferior to that of the best rung-4 DFT functional, ωB97M-V. By increasing the basis set for the MP2 part to def2-QZVPPD, we were able to substantially improve performance at modest cost (if an RI-MP2 approximation is made), with WTMAD2 for this G4(MP2)-XK-D method now comparable to the better rung-5 functionals (albeit at greater cost). A three-tier approach with a scaled MP3/def2-TZVPP intermediate step, however, leads to a G4(MP3)-D method that is markedly superior to even the best double hybrids ωB97M(2) and revDSD-PBEP86-D4. Evaluating the CCSD(T) component with a triple-ζ, rather than split-valence, basis set yields only a modest further improvement that is incommensurate with the drastic increase in computational cost. G4(MP3)-D and G4(MP2)-XK-D have about 40% better WTMAD2, at similar or lower computational cost, than their counterparts G4 and G4(MP2), respectively: detailed comparison reveals that the difference lies in larger molecules due to basis set incompleteness error. An E2/{T,Q} extrapolation and a CCSD(T)/def2-TZVP step provided the G4-T method of high accuracy and with just three fitted parameters. Using KS orbitals in MP2 leads to the G4(MP3|KS)-D method, which entirely eliminates the CCSD(T) step and has no steps costlier than scaled MP3; this shows a path forward to further improvements in double-hybrid density functional methods. None of our final selections require an empirical HLC correction; this cuts the number of empirical parameters in half and avoids discontinuities on potential energy surfaces. G4-T-DLPNO, a variant in which post-MP2 corrections are evaluated at the DLPNO-CCSD(T) level, achieves nearly the accuracy of G4-T but is applicable to much larger systems.

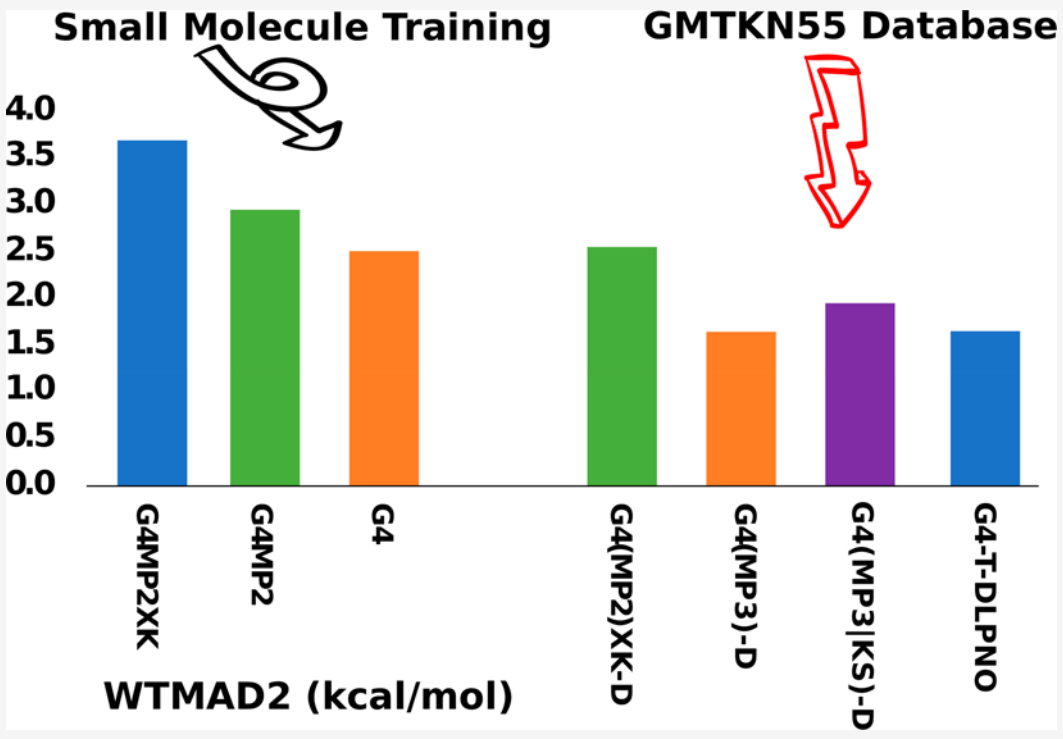

## I. INTRODUCTION

Among applied computational chemists, density functional theory (DFT) is presently the most widely used electronic structure approach, followed by wave function ab initio theory (WFT). WFT has a precisely defined Hamiltonian and a clear "road map" for systematic refinement, but in unmodified form hits an "exponential scaling wall" that limits application to small molecules. DFT "tunnels" through the scaling wall by reducing the many-electron Schrödinger equation to a set of coupled one-particle equations for an approximate exchange-correlation functional.[1] The accuracy of DFT stands or falls with the functional.

One well-established approach for reducing the computational cost of WFT methods has been the introduction of composite WFT (cWFT) protocols such as the following:

Gaussian-*n* theory (G*n*)[2−8] by the Pople group (see ref 9 for a review);

The CBS-QB3[10,11] and related methods[12] by Petersson and co-workers;

Multicoefficient correlation methods of Zhao and co-workers;[13−15]

In a higher accuracy regime, the ccCA approach[16−18] of Wilson and co-workers;

Table 1. Error Statistics (kcal/mol) for the GMTKN55 Database and Its Five Top-Level Subsets of Selected WFT, cWFT, and DFT Methods[a]

| methods | WTMAD2 | thermo | barrier | large | confor. | intermol. |
|---|---|---|---|---|---|---|
| | | | cWFT | | | |
| **G4(MP2)-XK-T** | 1.43 | 0.40 | 0.18 | 0.18 | 0.11 | 0.56 |
| **G4-Q-DLPNO** | 1.52 | 0.25 | 0.12 | 0.20 | 0.46 | 0.49 |
| **G4-T** | 1.51 | 0.35 | 0.16 | 0.23 | 0.14 | 0.63 |
| G4-T-DLPNO | 1.66 | 0.26 | 0.12 | 0.24 | 0.52 | 0.52 |
| **G4(MP3)-D** | 1.65 | 0.37 | 0.17 | 0.28 | 0.30 | 0.55 |
| **G4(MP3\|KS)-D** | 1.96 | 0.41 | 0.28 | 0.26 | 0.45 | 0.56 |
| **G4(MP2)-XK-D** | 2.56 | 0.46 | 0.29 | 0.34 | 0.68 | 0.79 |
| G4[4] | 2.52 | 0.38 | 0.23 | 0.75 | 0.38 | 0.78 |
| G4(MP2)[5] | 2.96 | 0.53 | 0.34 | 0.91 | 0.33 | 0.85 |
| CBS-QB3[10,11] | 3.10 | 0.40 | 0.35 | 0.60 | 0.20 | 1.55 |
| rev-G4MP2XK | 3.53 | 0.50 | 0.29 | 0.61 | 1.16 | 0.96 |
| MP2.X-Q | 3.29 | 0.71 | 0.78 | 0.88 | 0.42 | 0.50 |
| G4(MP2)-XK[53] | 3.71 | 0.45 | 0.31 | 0.67 | 1.25 | 1.02 |
| | | | WFT | | | |
| SCS-MP2-D3[89,b] | 5.22 | 1.23 | 0.95 | 1.39 | 0.91 | 0.75 |
| SCS-MP2[89] | 5.35 | 0.94 | 1.01 | 1.15 | 1.02 | 1.23 |
| MP2-D3[b] | 5.83 | 1.21 | 1.21 | 1.66 | 0.87 | 0.87 |
| MP2-D3[c] | 5.54 | 1.20 | 1.18 | 1.52 | 0.80 | 0.84 |
| MP2 | 6.91 | 1.21 | 1.23 | 1.78 | 1.47 | 1.21 |
| HF-D3[d] | 13.08 | 5.05 | 2.65 | 2.06 | 1.85 | 1.48 |
| HF | 29.46 | 5.87 | 3.74 | 3.66 | 7.27 | 8.92 |
| | | | DFT | | | |
| ωB97M(2)[49] | 2.19 | 0.44 | 0.26 | 0.42 | 0.58 | 0.49 |
| xrevDSD-PBEP86-D4[45] | 2.26 | 0.56 | 0.27 | 0.52 | 0.43 | 0.47 |
| revDSD-PBEP86-D4[45] | 2.33 | 0.56 | 0.31 | 0.58 | 0.41 | 0.48 |
| revDOD-PBEP86-D4[45] | 2.36 | 0.59 | 0.30 | 0.59 | 0.41 | 0.47 |
| revDSD-PBEP86-NL | 2.44 | 0.55 | 0.30 | 0.55 | 0.47 | 0.57 |
| revDSD-PBE-D4[45] | 2.46 | 0.65 | 0.35 | 0.53 | 0.43 | 0.50 |
| revDSD-PBEP86-D3[45] | 2.42 | 0.54 | 0.31 | 0.55 | 0.46 | 0.57 |
| revDSD-BLYP-D4[45] | 2.59 | 0.57 | 0.34 | 0.58 | 0.48 | 0.62 |
| DSD-SCAN-D4[45] | 2.64 | 0.60 | 0.40 | 0.62 | 0.45 | 0.56 |
| DSD-PBE-D4[42] | 2.64 | 0.61 | 0.39 | 0.56 | 0.53 | 0.54 |
| DSD-PBEP86-D4[42] | 2.65 | 0.54 | 0.37 | 0.63 | 0.55 | 0.56 |
| revDSD-PBEB95-D4[45] | 2.70 | 0.64 | 0.31 | 0.45 | 0.78 | 0.52 |
| DSD-BLYP-D4[45] | 2.83 | 0.58 | 0.38 | 0.59 | 0.68 | 0.60 |
| DSD-PBEP86-D3[42] | 3.10 | 0.55 | 0.45 | 0.49 | 0.65 | 0.97 |
| DSD-PBE-D3[42] | 3.17 | 0.66 | 0.41 | 0.54 | 0.73 | 0.83 |
| B2GP-PLYP-D3[90] | 3.19 | 0.63 | 0.42 | 0.66 | 0.64 | 0.85 |
| ωB97M-V[91] | 3.29 | 0.73 | 0.45 | 0.64 | 0.90 | 0.57 |
| ωB97X-V[92] | 3.96 | 1.02 | 0.56 | 1.07 | 0.73 | 0.58 |
| M06-2X-D3(0)[93] | 4.79 | 0.86 | 0.48 | 1.08 | 1.22 | 1.14 |
| B3LYP-D3 | 6.50 | 1.31 | 1.14 | 1.66 | 1.15 | 1.24 |

[a]D3 stands for D3(BJ) throughout this table; M06-2X was evaluated with a D3(0) correction, for want of D3BJ parameters. Statistics for M06-2X without D3(0) are essentially identical. Tabulated data for the DFT methods employing the def2-QZVPP basis set (def2-QZVPPD for subsets AHB21, G21EA, IL16, RG18, and WATER27) were obtained from refs 45 and 46 while the WFT (MP2, SCS-MP2, and HF) data in the same basis sets were obtained in this work, as were all cWFT results. The abbreviations “thermo”, “barrier”, “large”, “confor.”, and ‘intermol.’ stand for basic thermochemistry, barrier heights, reactions involving large molecules, conformer energies, and intermolecular interactions, respectively. [b]D3(BJ) parameters taken from Table S1 of ref 94. [c]$a_1 = 0$, $a_2 = 5.5$, $s_6 = -0.345$, $s_8 = 0$ (this work). [d]Present work; D3(BJ) parameters from Table 2 of the original D3(BJ) paper.[70]

In still higher accuracy regimes, the Weizmann-*n* approaches[19−24] of the present group and lower-cost modifications thereof (e.g., refs 25−27); the HEAT-n approaches of an international team around Stanton;[28−31] and the more general Feller−Peterson−Dixon (FPD) approach (refs 32−34 and references therein).

Especially the first two types of methods, being built into the popular Gaussian program system,[35] are applicable in a fairly black-box fashion, and hence are in fairly widespread use by nonspecialists. For recent reviews of cWFT methods and their performance for the thermochemistry and thermochemical kinetics of organic molecules, see Karton[36] and Chan;[37] for a recent general review of both experimental and computational thermochemistry, see Ruscic and Bross.[38]

One feature that cWFTs all share is the additivity approximation of the following form:

$$\text{HIGH/LARGE} = \text{HIGH/SMALL} + \text{LOW/LARGE} - \text{LOW/SMALL} + \text{Coupling}$$

$$\text{Coupling} = [\text{HIGH} - \text{LOW}]/\text{LARGE} - [\text{HIGH} - \text{LOW}]/\text{SMALL} \approx 0$$

in which LARGE and SMALL represent two different basis set sizes, HIGH and LOW represent a more rigorous and a more economical electron correlation level, respectively, and the cost savings derives from assuming the coupling term is negligible. (For a discussion of one-particle basis set−*n*-particle space coupling, see ref 39.)

Such approximations can be nested, e.g., in G4 theory,[4] which features three levels of electron correlation (MP2, MP4, and CCSD(T)).

While the more accurate of these methods, such as HEAT-*n*, Weizmann-*n*, and ccCA, achieve their performance without fitted empirical parameters, computational demands of even the most economical among them preclude routine use for larger molecules: for both ccCA and W1, CCSD(T) calculations in the spdf basis sets are required, and W1 requires CCSD/spdfg calculations, while ccCA “makes do” with MP2 calculations for the basis set extension steps. Our present focus will be on more economical methods similar in cost to G4(MP2) or G4 theory, i.e. (with one exception), requiring no basis sets larger than the polarized split-valence for the CCSD(T) step.

An approach that, in practical operation, combines elements of WFT and DFT methods is the double-hybrid density functional method (DHDFT,[40−45] see ref 46 for a recent review), which occupies the fifth rung on Perdew’s “Jacob’s Ladder”.[47] In DHDFT, both the exchange and the correlation terms are mixtures of DFT and WFT approaches (evaluated in a basis of Kohn−Sham orbitals). Using the very large GMTKN55 (general main-group thermochemistry, kinetics, and noncovalent interactions, 55 subsets) benchmark suite,[48] with nearly 2500 main-group molecules, we found that the best DHDFT functionals, ωB97M(2) by Mardirossian and Head-Gordon[49] and revDSD-PBEP86-D4 by our group,[45] have WTMAD2 (weighted mean absolute deviation) statistics around 2.2 kcal/mol, competitive with or superior to the cWFT methods we tested. Needless to say, double hybrids are computationally much more economical, especially if RI (resolution of the identity) approximations[50−52] are applied.

An additional impetus for the present work was the recent publication of the G4(MP2)-XK method,[53] which employs Weigend−Ahlrichs[54] def2 basis sets rather than Pople basis sets, and is thus applicable to the first five rows of the Periodic Table (H−Rn). Somewhat to our surprise, its WTMAD2 proved inferior to the best DHDFT functionals.

Deeper inspection of performance for the subsets revealed that, while the cWFT methods yielded better performance for heats of formation of small molecules (i.e., the W4-11 benchmark[55]), this was outweighed by degraded performance for larger-molecule subsets (see below and Table 1). As essentially all cDFT methods are parametrized against small-molecule training sets, this prompts the question whether this result was not an artifact of parametrization bias—if overall performance hadn't been "sacrificed on the altar of small-molecule thermochemistry", so to speak.

The purpose of the present paper is twofold. The first aim is to investigate whether superior, or simply more transferable, cWFT methods can be obtained by employing a large and diverse training set like GMTKN55. We shall show below that this is the case and shall present three options of ascending accuracy and computational cost.

The second objective is to see if such methods still have an edge in accuracy over the best DHDFT methods. We shall show that the answer is "yes, but not as big as you might expect".

## II. COMPUTATIONAL DETAILS

All calculations were carried out on the ChemFarm HPC cluster of the Faculty of Chemistry at the Weizmann Institute. CCSD(T) calculations,[56,57] as well as standard G4,[4] G4-(MP2),[5] and CBS-QB3[10,11] composite methods calculations, were performed using Gaussian 16, revision C.01;[35] medium-basis set MP3 and large-basis set MP2 (the latter using the RI-MP2 approximation[50,52]) calculations were performed using Q-CHEM 5.2;[58] the DLPNO-MP2[59] and DLPNO-CCSD-(T)[60] calculations were carried out using ORCA 4.2.1.[61] The conventional calculations benefited from 4TB SSD (solid state disk) arrays on our dedicated nodes, although some of the largest MP3 calculations exceeded that limit; for them we resorted to a 40TB shared-over-InfiniBand storage server custom-developed for us by Access Technologies of Ness Ziona, Israel.

As regards basis sets, for the standard G4, G4(MP2), and CBS-QB3 methods, the Pople basis sets specified in the original protocols were used without modification. Otherwise, we relied on the Weigend−Ahlrichs/Karlsruhe def2 family,[54] namely, the original def2-SVP, def2-TZVP, def2-TZVPP, and def2-QZVPP basis sets, as well as their diffuse function-augmented variants def2-SVPD, def2-TZVPPD, and def2-QZVPPD.[62] The unaugmented variants are available for all elements H−Rn, and the augmented variants for H−La and Hf−Rn.

For the original G4(MP2)-XK approach,[53] we employed the "minimally augmented" Karlsruhe basis sets ma-TZVP, ma-QZVP, ma-TZVXP (the prefix indicates addition of a single shell of diffuse valence exponents obtained by dividing the smallest s and p exponents by a factor of 3). In addition, the CCSD(T) step was carried out in a def2-SVSP basis set, which corresponds to def2-SVP with the polarization functions on hydrogen removed. Calculations were performed and results processed with the script provided in the Supporting Information to ref 53.

Where not already available internally in the respective codes, basis sets were retrieved from the Basis Set Exchange.[63] Auxiliary basis sets for RI-MP2 and DF-HF were used as stored in the Q-CHEM internal library; see refs 52, 64, 65, and 66 for the original references.

As in the original G4(MP2)-XK approach and the standard G4 and CBS-QB3 methods, all open-shell species were treated using UHF (unrestricted HF) orbitals. We mention in passing the existence of ROHF-based variants of G4(MP2) that are more suitable to radical research.[67]

For the alkaline and alkaline earth metals heavier than Ne, we correlated the $(n-1)$sp subvalence electrons, as it is well-known that these orbitals intrude into the valence band, especially of {O, F, Cl} compounds.[45] For the heavy p-block elements, we additionally correlated the $(n-1)$d subvalence electrons for similar reasons. As standard in def2 basis sets, small-core energy-consistent relativistic pseudopotentials[68] were used for elements heavier than Kr.

Dispersion corrections were evaluated using the Grimme et al. D3 model[69] with the Becke−Johnson damping function.[70] This combination is conventionally denoted D3(BJ). Based on our experience with double-hybrid functional parametrization,[45] the damping function's shape parameters were held fixed at $a_1 = s_8 = 0$, $a_2 = 5.5$, leaving the $R^{-6}$ overall scaling parameter $s_6$ as the only adjustable parameter.

For the DLPNO-CCSD(T), the "TightPNO" option[71] and the aforementioned settings of frozen inner orbitals were applied along with the original ($T_0$) triples approximation;[60] we used the "VeryTightSCF" convergence criteria and the RIJCOX approximation for constructing the Fock matrices.[72] Similarly, we used the def2-SVPD, def2-TZVPP, and def2-QZVPP basis sets along with the auxiliary versions of def2/J (see ref 73) and def2-SVP/C, def2-TZVPP/C, and def2-QZVPP/C (see ref 74) as implemented in ORCA. The core electrons were described by the def2 effective core potential (def2-ECP).[75] For the subsets AHB21, G21EA, IL16, RG18, and WATER27, we employed the diffuse-function augmented def2-TZVPPD and def2-QZVPPD,[62] inspired by ref 45. The DLPNO-CCSD(T)- based cWFT methods presented here have their energy difference CCSD-MP2 obtained from separate DLPNO-MP2 calculations rather than from the "semi-local (SL) MP2" energy obtained as a byproduct of the DLPNO-CCSD(T) step.

Our training set was the GMTKN55 benchmark;[48] as in our previous study on the revDSD functionals,[45] reference energetics, geometries, and charge/multiplicity information were obtained via the ACCDB database of Morgante and Peverati[76] and used as-is. (For the G4, G4(MP2), and CBS-QB3 implementations in Gaussian, "NoOpt" was specified to suppress the geometry optimization at these composite methods' standard levels of theory.) The calculations for the subsets C60ISO (isomerization energies of $C_{60}$ molecules)[77] and UPU23 (relative energies of uracil dinucleotides)[78] were currently not within reach for MP3 and canonical CCSD(T) methods. While the reported WTMAD2 values of DH-DFT and MP2-based WFT methods, i.e., SCS-MP2, are based on analysis of all subsets, excluding C60ISO and UPU23 does not appreciably affect WTMAD2 owing to the small (UPU23) and very small (C60ISO) weights these two subsets have in eq 1. For instance, C60ISO and UPU23 contribute only 0.3% and 2.5%, respectively, to the WTMAD2 for the ωB97M(2) double hybrid. Hence, despite excluding these two subsets, our WTMAD2 results are *de facto* equivalent to all-GMTKN55 WTMAD2 values, and we shall hence refer to them as GMTKN55 throughout. Statistical analysis was automated using a Fortran code developed in-house and available upon request. (We note that, as the reference data are complete basis set limit CCSD(T) calculations in the hypothetical motionless state, zero-point and thermal corrections need not be evaluated.)

The main performance metric for the GMTKN55 database is the WTMAD2,[48] which takes into account the different sizes and energy ranges of each subset; it is obtained according to the following equation:

$$\text{WTMAD2} = \frac{\sum_i^{55} \frac{56.84\ \text{kcal/mol}\cdot N_i\cdot \text{MAD}_i}{|\overline{\Delta E}|_i}}{\sum_i^{55} N_i} \quad (1)$$

where $\Delta\overline{E}$ corresponds to the mean signed deviation from the reference reaction energies for the $i$th subset, $N_i$ is the number of systems in the subset, and the $\text{MAD}_i$ is the mean absolute deviation between calculated and reference energies for the $i$th subset.

At one extreme, interaction energies $|\Delta E|$ for the RG18 subset[48] of rare-gas complexes range from 0.1 to 1.5 kcal/mol. At the other extreme, decomposition reaction energies $|\Delta E|$ for the MB16-43 subset[48] of artificial molecules run the gamut from −363 to 1290 kcal/mol. Thus, while a deviation of a few kcal/mol for an MB16-43 reaction would not materially affect the overall statistical error, a deviation of 1 kcal/mol at the RG18 subset would make it "shoot" up. Through trial and error, the application of numerous statistical schemes across the GMTKN24,[79] GMTKN30,[80] and GMTKN55[48] databases demonstrated in these papers that the balanced metric should encompass the number of reactions per subset $N_i$, the mean absolute deviation $\text{MAD}_i$ between the calculated and reference relative energies, and the total mean absolute deviation per subset $|\Delta\overline{E}|_i$. That metric was dubbed the weighted mean absolute deviation, type 2, conventionally abbreviated WTMAD2.[48] In prior papers[45,46,48] (as in the present work), MAD was favored over the RMSD (root-mean-square deviation) as MAD is a more "robust" measure[81] in the statistical sense of being less sensitive to one or a few outlier data points. For a pure normal distribution without systematic error, it can easily be shown[82] that $\frac{\text{RMSD}}{\text{MAD}} = \sqrt{\frac{\pi}{2}} = 1.25331 \ldots \approx \frac{5}{4}$, which may be helpful to readers more attuned to RMSD error measures. The MAD, RMSD, and the MAD/RMSD ratio values for each subset are included in the Supporting Information for the final selected methods.

The optimization of the parameters for each method was accomplished by the late Michael Powell's BOBYQA[83] (Bound Optimization BY Quadratic Approximation) derivative-free constrained optimizer. Different initial guesses were provided to ensure a global minimum was obtained for each optimization.

The reference energies of the GMTKN55 database are obtained from high-level ab initio methods including CCSD(T)/CBS, CCSD(T)-F12/CBS, or Weizmann-*n* theories such as W1-F12 or W2-F12.[84] These energies are nonrelativistic and ZPVE exclusive where all electrons were correlated for most calculations. For example, the references energies for the subset of rare-gas complexes RG18 were calculated at CCSD(T)/CBS(aug-cc-pVTZ/aug-cc-pVQZ), while for the subset of protonated isomers "PAREL", they were calculated at CCSD(T)/CBS(def2-TZVPP/def2-QZVPP).[48] The subsets of GMTKN55 are described in ref 48 as well as in the Supporting Information of the present paper.

## III. DESCRIPTION AND NOMENCLATURE OF THE THEORETICAL METHODS

**III.A. Description of the Theoretical Methods.** The original G4(MP2)-XK method has the following energy expression:

$$E = E_{\text{HF/CBS}} + \Delta E_{\text{SCS-MP2}} + \Delta E_{\text{scal-CCSD}} + \Delta E_{\text{scal-(T)}} + \text{HLC} \quad (2)$$

where each energy term is given by the following equations. The Hartree−Fock energy extrapolated to the complete basis set limit may be written as

$$E_{\text{HF/CBS}} = \frac{E_{\text{HF/ma-QZVP}} - E_{\text{HF/ma-TZVP}}\exp(-1.63)}{1 - \exp(-1.63)} \quad (3)$$

The correction terms for the spin-component-scaled MP2 correlation energy, $\Delta E_{\text{SCS-MP2}}$, the scaled coupled cluster singles−double correlation energy, and the coupled cluster triples excitations, are

$$\Delta E_{\text{SCS-MP2}} = c_3 E_{\text{MP2,OS/ma-TZVXP}} + c_4 E_{\text{MP2,SS/ma-TZVXP}} \quad (4)$$

$$\Delta E_{\text{scal-CCSD}} = c_5 E_{\text{CCSD/def2-SVSP}} - c_1 E_{\text{MP2,OS/def2-SVSP}} - c_2 E_{\text{MP2,SS/def2-SVSP}} \quad (5)$$

$$E_{\text{scal-(T)}} = c_6 E_{\text{C,(T)/def2-SVSP}} \quad (6)$$

The "high-level correction" HLC, like in G4 and G4(MP2) theory, is defined as

$$\text{HCL} = \begin{cases} -An_\beta & \text{for closed-shell molecules} \\ -A'n_\beta - B(n_\alpha - n_\beta) & \text{open-shell molecules} \\ -Cn_\beta - D(n_\alpha - n_\beta) & \text{for atomic species} \\ -En_\beta & \text{for "single-electron pair" species, such as } \text{Li}_2 \end{cases} \quad (7)$$

The variables $n_\alpha$ and $n_\beta$ (by convention $n_\alpha \geq n_\beta$) correspond to the number of $\alpha$ and $\beta$ valence electrons, respectively, according to the conventional largest noble-gas-core definition. The parameters $A$, $A'$, $B$, $C$, $D$, and $E$ were established through parameter optimization based on the chosen training set of reference energies. The HLC was originally incorporated in the Gaussian-*n* theories and the recent G4(MP2)-XK method to account for residual basis set incompleteness error; its oldest incarnation was based on correcting the total energies for the hydrogen atom and the hydrogen molecule.[7]

The inexpensive two-tier methods considered here share the G4(MP2)-XK energy expression, albeit with different parameters.

The three-tier methods considered have an energy expression of the following energy form, e.g., when Schwenke-type[85,86] MP2 extrapolation is used:

$$E = E_{\mathrm{HF/CBS}} + \Delta E_{\mathrm{MP2+MP3}} + \Delta E_{\mathrm{scal\text{-}CCSD}} + \Delta E_{\mathrm{scal\text{-}(T)}} + \mathrm{HLC} + E_{\mathrm{disp}} \tag{8}$$

The first term is the extrapolated Hartree−Fock energy of the form

$$E_{\mathrm{HF/CBS}} = \frac{E_{\mathrm{HF/def2\text{-}QZVPPD}} - E_{\mathrm{HF/def2\text{-}TZVPPD}}\exp(-1.63)}{1 - \exp(-1.63)} \tag{9}$$

The remaining terms are the post-HF correlation corrections. The energy difference, $\Delta E_{\mathrm{MP2+MP3}}$, combining the third-order MP correction and the second-order correlation term at the extrapolation limit is given by

$$\Delta E_{\mathrm{MP2+MP3}} = c_2 E_{\mathrm{[MP3\text{-}MP2]/def2\text{-}TZVPP}} - [(c_1 + 1)E_{\mathrm{corr,MP2/def2\text{-}QZVPPD}} - c_1 E_{\mathrm{corr,MP2/def2\text{-}TZVPPD}}] \tag{10}$$

The post-MP3 improvements to the total energy from the coupled cluster wave function are

$$\Delta E_{\mathrm{C,CCSD\text{-}MP3}} = c_3(E_{\mathrm{C,CCSD/def2\text{-}SVSP}} - E_{\mathrm{MP3/def2\text{-}SVSP}}) \tag{11}$$

$$\Delta E_{\mathrm{scal\text{-}(T)}} = c_4 E_{\mathrm{C,(T)/def2\text{-}SVSP}} \tag{12}$$

The last two terms involve the empirical dispersion correction and the HLC.

$$E_{\mathrm{disp}} = c_5 E[\mathrm{D3(BJ)}] \tag{13}$$

Finally, in the most expensive two-tier method we considered, CCSD(T)/def2-TZVP is the smallest basis set step, which makes the separate MP3/def2-TZVPP step redundant. Hence, we have

$$E = E_{\mathrm{HF/CBS}} + \Delta E_{\mathrm{SCS\text{-}MP2}} + \Delta E_{\mathrm{C,CCSD\text{-}MP2}} + \Delta E_{\mathrm{scal\text{-}(T)}} + \mathrm{HLC} \tag{14}$$

The Hartree−Fock extrapolated energy, $E_{\mathrm{HF/CBS}}$, is likewise given by eq 9. The post-HF energy corrections arise from MP2, CCSD(T), and the dispersion correction. For the avoidance of doubt, the coefficients in the SCS-MP2 term are adjustable parameters and not standard[87,88] SCS-MP2 coefficients.

$$\Delta E_{\mathrm{SCS\text{-}MP2}} = c_1 E_{\mathrm{MP2,OS/def2\text{-}QZVPPD}} + c_2 E_{\mathrm{MP2,SS/def2\text{-}QZVPPD}} \tag{15}$$

The remaining corrections are the energy difference between CCSD and MP2 in the same basis set, $\Delta E_{\mathrm{c,CCSD\text{-}MP2}}$, the triples excitations correlation energy, and the dispersion energy:

$$\Delta E_{\mathrm{C,CCSD\text{-}MP2}} = c_3[E_{\mathrm{CCSD/def2\text{-}TZVP}} - E_{\mathrm{MP2/def2\text{-}TZVP}}] \tag{16}$$

$$\Delta E_{\mathrm{scal\text{-}(T)}} = c_4 E_{\mathrm{C,(T)/def2\text{-}TZVP}} \tag{17}$$

$$E_{\mathrm{disp}} = c_5 E[\mathrm{D3(BJ)}] \tag{18}$$

In all of the above, varying the HF extrapolation parameter (originally from ref 4) had only insignificant effect, and we have hence held it fixed. Detailed equations and parameters for all individual composite methods considered in this work are given in the Supporting Information.

A minor but significant practical detail about the HLC needs to be clarified here. When introducing separate coefficients for molecules and separate atoms in the HLC, the authors of the original G4 papers probably never envisaged an application to noble gas complexes: hence, the unmodified HLC yields an abnormally large relative error in the RG18 subset, which through RG18's large weight in WTMAD2 has an effect of several kcal/mol there. The problem can be solved by the simple expedient of treating the closed-shell rare gas atoms as if they were molecules: we have done so throughout the present paper. The original G4(MP2)-XK yields a discouraging WTMAD2 of 6.31 kcal/mol; a closer inspection revealed a contribution of 1.58 kcal/mol to the WTMAD2 from the RG18 subset. After treating the "rare gas atoms" as molecules, the RG18 contribution to WTMAD2 plummets to 0.17 kcal/mol, bringing the total down to 3.71 kcal/mol.

**III.B. Nomenclature of the Theoretical Methods.** In this section, we focus on the nomenclature of the theoretical methods, providing concise and unambiguous names. The naming scheme is illustrated in Figure 1. There are two different ways of naming the methods based on the E2 extrapolation: either G4-scs-*n* and G4(scsMP2.X)-*n*, if the same-spin and opposite-spin E2 components are separately extrapolated, or G4-*n* and G4(MP2.X)-*n*, if the total E2 energy is extrapolated. The presence of the MP3/def2-TZVPP step is denoted as "MP2.X". We will use the suffix KS if the MP2(-like) term is evaluated in a basis of Kohn−Sham orbitals with the largest basis set; additionally including an MP3/def2-TZVPP step determines the "MP3" label. The DLPNO suffix indicates the use of DLPNO-CCSD(T) instead of canonical CCSD(T). The basis set for the canonical CCSD(T) step is noted in abbreviated form as "D" or 'T" for def2-SVSP or def2-TZVP, respectively. Similarly, for the DLPNO-CCSD(T), the basis set is denoted as "D", "T", or "Q" for def2-SVPD, def2-TZVPP, or def2-QZVPP, respectively.

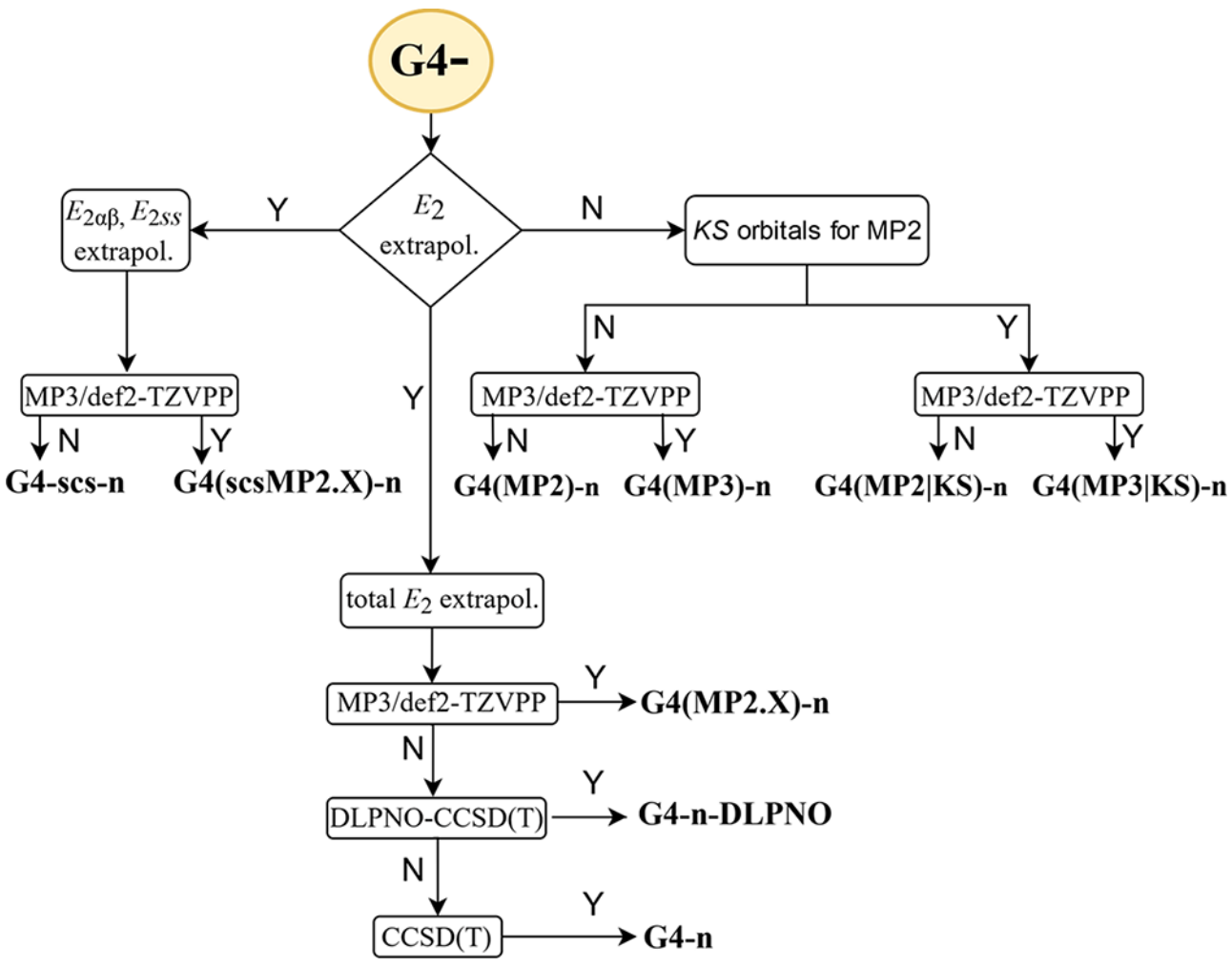

**Figure 1.** Nomenclature scheme of the theoretical methods.

**Table 2. WTMAD2 Values (kcal/mol) for Selected Standard cWFT and Presently Developed cWFT Methods, as Well as Optimized Parameters for the Latter**

| | | | | | | coefficients of the energy components | | | | | | | HLC coefficients | | | | | |
|---|---|---|---|---|---|---|---|---|---|---|---|---|---|---|---|---|---|---|
| MP2 | MP3 | CCSD(T) | $N_{param}$ | method | WTMAD2 | $c_{(E2,os)'}$ | $c_{(E2,ss)'}$ | $c_{E2,os}$ | $c_{E2,ss}$ | $c_{E(C,CCSD)}$ | $c_T$ | $c_{Disp}$ | $A$ | $A'$ | $B$ | $C$ | $D$ | $E$ |
| T | — | D | 12 | G4(MP2)-XK | 3.71 | 1.131 | 0.512 | 1.041 | 0.704 | 1.048 | 0.526 | [0] | 9.369 | 9.449 | 3.832 | 9.594 | 1.874 | 2.491 |
| T | — | D | 12 | revG4(MP2)-XK-H6-v1 | 3.53 | 1.307 | 0.385 | 1.170 | 0.614 | 0.984 | 0.736 | [0] | 8.540 | 8.603 | 2.199 | 7.771 | 1.052 | 3.108 |
| Q | — | D | 12 | G4(MP2)-XK-D-H5-v1 | 2.53 | 1.216 | 0.660 | 1.008 | 0.925 | 1.062 | 0.778 | −0.276 | 7.082 | 7.082 | 2.263 | 7.198 | 1.379 | 3.292 |
| Q | — | D | 11 | G4(MP2)-XK-D-H5-v2 | 2.64 | 1.270 | 0.417 | 1.069 | 0.669 | 1.041 | 0.701 | [0] | 7.262 | 7.262 | 2.035 | 6.999 | 1.420 | 3.467 |
| Q | — | D | 13 | G4(MP2)-XK-D-H6-v1 | 2.52 | 1.210 | 0.616 | 1.026 | 0.865 | 1.042 | 0.825 | −0.227 | 7.161 | 7.197 | 2.081 | 7.131 | 1.584 | 3.546 |
| Q | — | D | 7 | **G4(MP2)-XK-D-v1** | 2.56 | 1.309 | 0.674 | 1.124 | 0.890 | 1.113 | 0.699 | −0.383 | [0] | [0] | [0] | [0] | [0] | [0] |
| Q | — | D | 6 | G4(MP2)-XK-D-v2 | 2.73 | 1.482 | 0.271 | 1.320 | 0.457 | 1.052 | 0.778 | [0] | [0] | [0] | [0] | [0] | [0] | [0] |
| Q | — | T | 13 | G4(MP2)-XK-T-H6-v1 | 1.36 | 1.227 | 1.030 | 1.128 | 1.128 | 1.029 | 1.000 | −0.095 | 7.139 | 7.172 | 1.886 | 7.262 | 1.591 | 3.513 |
| Q | — | T | 6 | **G4(MP2)-XK-T-v2** | 1.43 | 1.609 | 0.720 | 1.516 | 0.722 | 1.068 | 1.051 | [0] | [0] | [0] | [0] | [0] | [0] | [0] |
| | | | | | | $c_{E2,ss}$ | $c_{E2,os}$ | $c_{CCSD-MP2}$ | $c_T$ | $c_{Disp}$ | | | $A$ | $A'$ | $B$ | $C$ | $D$ | $E$ |
| Q | — | D | 10 | G4(MP2)-D-H5-v1 | 2.59 | 1.224 | 0.865 | 1.038 | 0.881 | −0.294 | | | 7.456 | 7.456 | 2.268 | 7.318 | 1.043 | 3.653 |
| Q | — | D | 9 | G4(MP2)-D-H5-v2 | 2.71 | 1.106 | 0.852 | 0.956 | 0.837 | [0] | | | 7.547 | 7.547 | 2.206 | 7.385 | 0.278 | 4.005 |
| Q | — | D | 5 | G4(MP2)-D-v1 | 2.68 | 1.226 | 0.977 | 1.111 | 0.776 | −0.456 | | | [0] | [0] | [0] | [0] | [0] | [0] |
| Q | — | D | 4 | G4(MP2)-D-v2 | 3.01 | 1.045 | 0.968 | 1.033 | 0.796 | [0] | | | [0] | [0] | [0] | [0] | [0] | [0] |
| | | | | | | $c_{(E2/CBS)}$ | $c_{CCSD-MP2}$ | $c_{T_{cT}}$ | $c_{Disp}$ | | | | $A$ | $A'$ | $B$ | $C$ | $D$ | $E$ |
| {T, Q} | — | T | 10 | G4-T-H6-v1 | 1.42 | 0.312 | 1.024 | 1.053 | −0.025 | | | | 7.046 | 7.107 | 2.783 | 6.880 | 2.598 | 4.298 |
| {T, Q} | — | T | 4 | G4-T-v1 | 1.48 | 0.550 | 1.063 | 1.187 | −0.070 | | | | [0] | [0] | [0] | [0] | [0] | [0] |
| {T, Q} | — | T | 3 | **G4-T-v2** | 1.51 | 0.542 | 1.060 | 1.114 | [0] | | | | [0] | [0] | [0] | [0] | [0] | [0] |
| {T, Q} | — | T(DLPNO) | 3 | **G4-T-DLPNO-v2** | 1.66 | 0.601 | 1.019 | 1.204 | [0] | | | | [0] | [0] | [0] | [0] | [0] | [0] |
| {T, Q} | — | Q(DLPNO) | 3 | **G4-Q-DLPNO-v2** | 1.52 | 0.513 | 1.003 | 1.185 | [0] | | | | [0] | [0] | [0] | [0] | [0] | [0] |
| | | | | | | $c_{E2,ss}$ | $c_{E2,os}$ | $c_{E3}$ | $c_{CCSD-MP3}$ | $c_T$ | $c_{Disp}$ | | $A$ | $A'$ | $B$ | $C$ | $D$ | $E$ |
| Q | T | D | 12 | G4(MP3)-D-H6-v1 | 1.63 | 1.250 | 0.951 | 1.033 | 1.039 | 0.566 | −0.087 | | 7.138 | 7.173 | 2.228 | 7.162 | 1.467 | 3.581 |
| Q | T | D | 6 | **G4(MP3)-D-v1** | 1.65 | 1.284 | 1.018 | 1.089 | 1.119 | 0.411 | −0.185 | | [0] | [0] | [0] | [0] | [0] | [0] |
| KSQ | T | D | 11 | G4(MP3\|KS)-D-H5-v1 | 1.89 | 0.831 | 0.944 | 0.601 | −0.089 | 0.057 | −0.056 | | 7.614 | 7.614 | 2.293 | 7.828 | 1.525 | 3.600 |
| KSQ | T | — | 9 | G4(MP3\|KS)-D-H5-v5 | 1.89 | 0.845 | 0.938 | 0.596 | [0] | [0] | −0.053 | | 7.505 | 7.505 | 2.057 | 7.639 | 1.544 | 3.861 |
| KSQ | T | — | 8 | G4(MP3\|KS)-D-H5-v6 | 1.89 | 0.822 | 0.938 | 0.566 | [0] | [0] | [0] | | 6.233 | 6.233 | 3.006 | 6.566 | 1.076 | 3.366 |
| KSQ | T | — | 4 | G4(MP3\|KS)-D-v5 | 1.93 | 0.795 | 1.013 | 0.623 | [0] | [0] | −0.099 | | [0] | [0] | [0] | [0] | [0] | [0] |
| KSQ | T | — | 3 | **G4(MP3\|KS)-D-v6** | 1.96 | 0.751 | 1.023 | 0.609 | [0] | [0] | [0] | | [0] | [0] | [0] | [0] | [0] | [0] |
| Q | T | — | 4 | MP2.X-Q | 3.28 | 1.007 | 1.128 | 0.816 | [0] | [0] | −0.033 | | [0] | [0] | [0] | [0] | [0] | [0] |
| {T, Q} | T | D | 4 | G4(scsMP2.X)-D-v8 | 1.84 | 0.633 | 0.906 | 0.965 | 1.026 | 1.026 | [0] | | [0] | [0] | [0] | [0] | [0] | [0] |
| | | | | G4[4] | 2.52 | | | | | | | | | | | | | |
| | | | | G4MP2[5] | 2.96 | | | | | | | | | | | | | |
| | | | | CBS-QB3[11] | 3.10 | | | | | | | | | | | | | |

| method | energy terms included |
|---|---|
| G4(MP2)-XK | $= E_{HF/CBS} + c_3 E_{MP2,OS/ma\text{-}TZVXP} + c_4 E_{MP2,SS/ma\text{-}TZVXP} + c_5 E_{CCSD/def2\text{-}SVSP} - c_1 E_{MP2,OS/def2\text{-}SVSP} - c_2 E_{MP2,SS/def2\text{-}SVSP} + c_6 E_{C,(T)/def2\text{-}SVSP} + HLC$ |
| revG4(MP2)-XK-H6-v1 | $= E_{HF/CBS} + c_3 E_{MP2,OS/ma\text{-}TZVXP} + c_4 E_{MP2,SS/ma\text{-}TZVXP} + c_5 E_{CCSD/def2\text{-}SVSP} - c_1 E_{MP2,OS/def2\text{-}SVSP} - c_2 E_{MP2,SS/def2\text{-}SVSP} + c_6 E_{C,(T)/def2\text{-}SVSP} + HLC$ |

Table 2. continued

| method | energy terms included |
| --- | --- |
| G4(MP2)-XK-D-H5-v1 | $= E_{HF/CBS} + c_3E_{MP2,OS/def2\text{-}QZVPPD} + c_4E_{MP2,SS/def2\text{-}QZVPPD}$ $+ c_5E_{CCSD/def2\text{-}SVSP} - c_1E_{MP2,OS/def2\text{-}SVSP} - c_2E_{MP2,SS/def2\text{-}SVSP}$ $+ c_6E_{C,(T)/def2\text{-}SVSP} + HLC' + c_7E[D3BJ]$ |
| G4(MP2)-XK-D-H5-v2 | $= E_{HF/CBS} + c_3E_{MP2,OS/def2\text{-}QZVPPD} + c_4E_{MP2,SS/def2\text{-}QZVPPD} + c_5E_{CCSD/def2\text{-}SVSP} - c_1E_{MP2,OS/def2\text{-}SVSP} - c_2E_{MP2,SS/def2\text{-}SVSP} + c_6E_{C,(T)/def2\text{-}SVSP} + HLC'$ |
| G4(MP2)-XK-D-H6-v1 | $= E_{HF/CBS} + c_3E_{MP2,OS/def2\text{-}QZVPPD} + c_4E_{MP2,SS/def2\text{-}QZVPPD}$ $+ c_5E_{CCSD/def2\text{-}SVSP} - c_1E_{MP2,OS/def2\text{-}SVSP} - c_2E_{MP2,SS/def2\text{-}SVSP}$ $+ c_6E_{C,(T)/def2\text{-}SVSP} + HLC + c_7E[D3BJ]$ |
| **G4(MP2)-XK-D-v1** | $= E_{HF/CBS} + c_3E_{MP2,OS/def2\text{-}QZVPPD} + c_4E_{MP2,SS/def2\text{-}QZVPPD}$ $+ c_5E_{CCSD/def2\text{-}SVSP} - c_1E_{MP2,OS/def2\text{-}SVSP} - c_2E_{MP2,SS/def2\text{-}SVSP}$ $+ c_6E_{C,(T)/def2\text{-}SVSP} + c_7E[D3BJ]$ |
| G4(MP2)-XK-D-v2 | $= E_{HF/CBS} + c_3E_{MP2,OS/def2\text{-}QZVPPD} + c_4E_{MP2,SS/def2\text{-}QZVPPD} + c_5E_{CCSD/def2\text{-}SVSP} - c_1E_{MP2,OS/def2\text{-}SVSP} - c_2E_{MP2,SS/def2\text{-}SVSP} + c_6E_{C,(T)/def2\text{-}SVSP}$ |
| G4(MP2)-XK-T-H6-v1 | $= E_{HF/CBS} + c_3E_{MP2,OS/def2\text{-}QZVPPD} + c_4E_{MP2,SS/def2\text{-}QZVPPD}$ $+ c_5E_{CCSD/def2\text{-}TZVP} - c_1 - E_{MP2,OS/def2\text{-}TZVP} - c_2E_{MP2,SS/def2\text{-}TZVP}$ $+ c_6E_{C,(T)/def2\text{-}TZVP} + HLC + c_7E[D3BJ]$ |
| **G4(MP2)-XK-T-v2** | $= E_{HF/CBS} + c_3E_{MP2,OS/def2\text{-}QZVPPD} + c_4E_{MP2,SS/def2\text{-}QZVPPD}$ $+ c_5E_{CCSD/def2\text{-}TZVP} - c_1E_{MP2,OS/def2\text{-}TZVP} - c_2E_{MP2,SS/def2\text{-}TZVP}$ $+ c_6E_{C,(T)/def2\text{-}TZVP}$ |
| G4(MP2)-D-H5-v1 | $= E_{HF/CBS} + c_1E_{MP2,OS/def2\text{-}QZVPPD} + c_2E_{MP2,SS/def2\text{-}QZVPPD}$ $+ c_3(E_{C,CCSD/def2\text{-}SVSP} - E_{MP2/def2\text{-}SVSP}) + c_4E_{C,(T)/def2\text{-}SVSP} + HLC'$ $+ c_5E[D3BJ]$ |
| G4(MP2)-D-H5-v2 | $= E_{HF/CBS} + c_1E_{MP2,OS/def2\text{-}QZVPPD} + c_2E_{MP2,SS/def2\text{-}QZVPPD}$ $+ c_3(E_{C,CCSD/def2\text{-}SVSP} - E_{MP2/def2\text{-}SVSP}) + c_4E_{C,(T)/def2\text{-}SVSP} + HLC'$ |
| G4(MP2)-D-v1 | $= E_{HF/CBS} + c_1E_{MP2,OS/def2\text{-}QZVPPD} + c_2E_{MP2,SS/def2\text{-}QZVPPD}$ $+ c_3(E_{C,CCSD/def2\text{-}SVSP} - E_{MP2/def2\text{-}SVSP}) + c_4E_{C,(T)/def2\text{-}SVSP} + c_5E[D3BJ]$ |
| G4(MP2)-D-v2 | $= E_{HF/CBS} + c_1E_{MP2,OS/def2\text{-}QZVPPD} + c_2E_{MP2,SS/def2\text{-}QZVPPD}$ $+ c_3(E_{C,CCSD/def2\text{-}SVSP} - E_{MP2/def2\text{-}SVSP}) + c_4E_{C,(T)/def2\text{-}SVSP}$ |
| G4-T-H6-v1 | $= E_{HF/CBS} + [(c_1 + 1)E_{corr,MP2/def2\text{-}QZVPPD} - c_1E_{corr,MP2/def2\text{-}TZVPPD}]$ $+ c_2(E_{C,CCSD/def2\text{-}TZVP}E_{MP2/def2\text{-}TZVP}) + c_3E_{C,(T)/def2\text{-}TZVP} + HLC$ $+ c_4E[D3BJ]$ |
| G4-T-v1 | $= E_{HF/CBS} + [(c_1 + 1)E_{corr,MP2/def2\text{-}QZVPPD} - c_1E_{corr,MP2/def2\text{-}TZVPPD}] + c_2(E_{C,CCSD/def2\text{-}TZVP} - E_{MP2/def2\text{-}TZVP}) + c_3E_{C,(T)/def2\text{-}TZVP} + c_4E[D3BJ]$ |

Table 2. continued

| method | energy terms included |
|---|---|
| **G4-T-v2** | $= E_{HF/CBS} + [(c_1 + 1)E_{corr,MP2/def2\text{-}QZVPPD} - c_1E_{corr,MP2/def2\text{-}TZVPPD}] + c_2(E_{C,CCSD/def2\text{-}TZVP} - E_{MP2/def2\text{-}TZVP}) + c_3E_{C,(T)/def2\text{-}TZVP}$ |
| **G4-T-DLPNO-v2** | $= E_{HF/CBS} + [(c_1 + 1)E_{corr,MP2/def2\text{-}QZVPPD} - c_1E_{corr,MP2/def2\text{-}TZVPPD}] + c_2(E_{C,DLPNO\text{-}CCSD/def2\text{-}TZVPP} - E_{C,(DLPNO\text{-}MP2)/def2\text{-}TZVPP}) + c_3E_{C,DLPNO\text{-}(T)/def2\text{-}TZVPP}$ |
| **G4-Q-DLPNO-v2** | $= E_{HF/CBS} + [(c_1 + 1)E_{corr,MP2/def2\text{-}QZVPPD} - c_1E_{corr,MP2/def2\text{-}TZVPPD}] + c_2(E_{C,DLPNO\text{-}CCSD/def2\text{-}QZVPP} - E_{C,(DLPNO\text{-}MP2)/def2\text{-}QZVPP}) + c_3E_{C,DLPNO\text{-}(T)/def2\text{-}QZVPP}$ |
| G4(MP3)-D-H6-v1 | $= E_{HF/CBS} + c_1E_{MP2,OS/def2\text{-}QZVPPD} + c_2E_{MP2,SS/def2\text{-}QZVPPD} + c_3E_{[MP3\text{-}MP2]/def2\text{-}TZVPP} + c_4E_{C,[CCSD\text{-}MP3]/def2\text{-}SVSP} + c_5E_{C,(T)/def2\text{-}SVSP} + \text{HLC} + c_6E[\text{D3BJ}]$ |
| **G4(MP3)-D-v1** | $= E_{HF/CBS} + c_1E_{MP2,OS/def2\text{-}QZVPPD} + c_2E_{MP2,SS/def2\text{-}QZVPPD} + c_3E_{[MP3\text{-}MP2]/def2\text{-}TZVPP} + c_4E_{C,[CCSD\text{-}MP3]/def2\text{-}SVSP} + c_5E_{C,(T)/def2\text{-}SVSP} + c_6E[\text{D3BJ}]$ |
| G4(MP3\|KS)-D-H5-v1 | $= E_{HF/CBS} + c_1E_{MP2\|KS,OS/def2\text{-}QZVPPD} + c_2E_{MP2\|KS,SS/def2\text{-}QZVPPD} + c_3E_{[MP3\text{-}MP2]/def2\text{-}TZVPP} + c_4E_{C,[CCSD\text{-}MP3]/def2\text{-}SVSP} + c_5E_{C,(T)/def2\text{-}SVSP} + \text{HLC}' + c_6E[\text{D3BJ}]$ |
| G4(MP3\|KS)-D-H5-v5 | $= E_{HF/CBS} + c_1E_{MP2\|KS,OS/def2\text{-}QZVPPD} + c_2E_{MP2\|KS,SS/def2\text{-}QZVPPD} + c_3E_{[MP3\text{-}MP2]/def2\text{-}TZVPP} + \text{HLC}' + c_6E[\text{D3BJ}]$ |
| G4(MP3\|KS)-D-H5-v6 | $= E_{HF/CBS} + c_1E_{MP2\|KS,OS/def2\text{-}QZVPPD} + c_2E_{MP2\|KS,SS/def2\text{-}QZVPPD} + c_3E_{[MP3\text{-}MP2]/def2\text{-}TZVPP} + \text{HLC}'$ |
| G4(MP3\|KS)-D-v5 | $= E_{HF/CBS} + c_1E_{MP2\|KS,OS/def2\text{-}QZVPPD} + c_2E_{MP2\|KS,SS/def2\text{-}QZVPPD} + c_3E_{[MP3\text{-}MP2]/def2\text{-}TZVPP} + c_6E[\text{D3BJ}]$ |
| **G4(MP3\|KS)-D-v6** | $= E_{HF/CBS} + c_1E_{MP2\|KS,OS/def2\text{-}QZVPPD} + c_2E_{MP2\|KS,SS/def2\text{-}QZVPPD} + c_3E_{[MP3\text{-}MP2]/def2\text{-}TZVPP}$ |
| MP2.X-Q | $= E_{HF/CBS} + c_1E_{MP2,OS/def2\text{-}QZVPPD} + c_2E_{MP2,SS/def2\text{-}QZVPPD} + c_3E_{[MP3\text{-}MP2]/def2\text{-}TZVPP}$ |

**Table 2. continued**

| method | energy terms included |
|---|---|
| G4(scsMP2.X)-D-v8 | $= E_{HF/CBS} + [(c_1 + 1)E_{C,SS\text{-}MP2/def2\text{-}QZVPPD} - c_1E_{C,SS-MP2/def2\text{-}TZVPPD}]$ $+ [(c_2 + 1)E_{C,OS-MP2/def2\text{-}QZVPPD} - c_2E_{C,OS-MP2/def2\text{-}TZVPPD}]$ $+ c_3E_{[MP3\text{-}MP2]/def2\text{-}TZVPP} + c_4E_{C,[CCSD\text{-}MP3]/def2\text{-}SVSP}$ $+ c_5E_{C,(T)/def2\text{-}SVSP}(c_4 = c_5)$ |
| G4[4] | $= [E_{HF/mod\text{-}aug\text{-}cc\text{-}pV\{Q,5\}Z} - E_{HF/G3LargeXP}] + [E_{MP2(full)/G3LargeXP}$ $- E_{MP2/6\text{-}31G(2df,p)} - E_{MP2/6\text{-}31+G(d)} + E_{MP2/6\text{-}31G(d)}] + E_{MP4/6\text{-}31G(d)}$ $+ [E_{MP4/6\text{-}31+G(d)} - E_{MP4/6\text{-}31G(d)}] + [E_{MP4/6\text{-}31G(2df,p)} - E_{MP4/6\text{-}31G(d)}]$ $+ [E_{CCSD(T)/6\text{-}31G(d)} - E_{MP4/6\text{-}31G(d)}] + HLC$ |
| G4MP2[5] | $= [E_{HF/mod\text{-}aug\text{-}cc\text{-}pV\{Q,5\}Z} - E_{HF/G3LargeXP}] + [E_{MP2(FC)/G3MP2LargeXP}$ $- E_{MP2(FC)/6\text{-}31G(d)}] + E_{CCSD(T,FC)/6\text{-}31G(d)} + HLC$ |
| CBS-QB3[11] | $= E_{MP2/6\text{-}311+G(3d2f,2df,2p)} + \Delta E(CBS^{(2)}) + \Delta E(spin)$ $+ E_{[MP4(SDQ)\text{-}MP2]/6-31+G(d(f),p)} + E_{[CCSD(T)\text{-}MP4(SDQ)]/6\text{-}31+G^{\dagger}}$ $+ \Delta E(CBS - int) + \Delta E(emp)$ |

[a]The HF{T,Q} extrapolation was used for all methods. The coefficients $c_{(E2,os)}$, and $c_{(E2,ss)}$, correspond to the $E_{2\alpha\beta}$ and $E_{2ss}$ components from the CCSD(T) step, respectively. The final methods are highlighted. All methods are contained in the full table in the Supporting Information and highlighted are the final selected methods.

Regarding the HLC, we add the suffix H6 to indicate the original HLC of G4(MP2)-XK. If the constraint $A = A'$ is imposed on eq 7, this is indicated by the H5 label, while the absence of either label indicates that the HLC was eliminated altogether. Overall, we tried numerous variants of the composite methods described herein, namely, 291 variants for the canonical CCSD(T)-based cWFTs, 18 variants for DLNPO-CCSD(T)-, and 18 variants for DLPNO-CCSD($T_1$)-based methods. Said variants differ from each other in the number of optimized parameters or HLC treatment; e.g., some parameters might be set equal to zero, such as the triples correction term, or some pairs of parameters might be set to equal during the parametrization. Therefore, we have adopted the suffix "v*x*", where *x* is the variant number. The equations of all composite methods are explicitly listed in the Supporting Information.

A few examples will further clarify the nomenclature. When we use the formalism of the original G4(MP2)-XK, the methods' names retain the "XK" label, and suffixes contain information about the HLC and the basis set used for the CCSD(T) step. G4(MP2)-XK-T-H6-v1 has an energy expression similar to the original G4(MP2)-XK; the letter "T" stands for the CCSD(T)/def2-TZVP step and "H6" for the six-parameter HLC, while "v1" denotes the variant with dispersion correction included. One primary goal was to limit the empirical parameters of the composite methods, and the "XK" label is omitted for the minimally empirical methods where we did not scale the E2 correlation energy of the smaller basis set compared to the original G4(MP2)-XK. G4-T-v2 is based on a CCSD(T)/def2-TZVP step, without HLC, and the suffix "v2" stands for the second variant of the G4-T. As we proceed toward the final choice of the minimally empirical methods, we will drop the variant label "vn" for the best overall methods.

## IV. RESULTS AND DISCUSSION

Full statistics and parameters for all the different empirical methods considered are given in the Supporting Information. The component breakdown of WTMAD2 for selected methods is presented in Table 1. A selection of the most relevant data is summarized by the rows in Table 2, and the final selected composite methods are briefly presented in Table 3. Figure 2 depicts WTMAD2 values for several DFT functionals and cWFT methods taken from refs 45 and 46 as well as for the cWFT methods developed in the present work.

**IV.A. Using Our Largest Available Basis Sets.** Our best result overall is WTMAD2 = 1.36 kcal/mol (G4(MP2)-XK-T-H6-v1, Table 2). Interestingly, replacing def2-QZVPPD by an extrapolation E2/{T,Q} does not improve WTMAD2 and in fact slightly raises it to 1.42 kcal/mol (G4-T-H6-v1).

**Table 3. Summary of Recommended cWFT Methods**

| method | WTMAD2 | parameters |
|---|---|---|
| G4(MP2)XK-T | 1.43 | 6 |
| G4-T | 1.51 | 3 |
| G4-Q-DLPNO | 1.52 | 3 |
| G4-T-DLPNO | 1.66 | 3 |
| G4(MP3)-D | 1.65 | 6 |
| G4(MP3\|KS)-D | 1.96 | 3 |
| G4(MP2)XK-D | 2.56 | 7 |

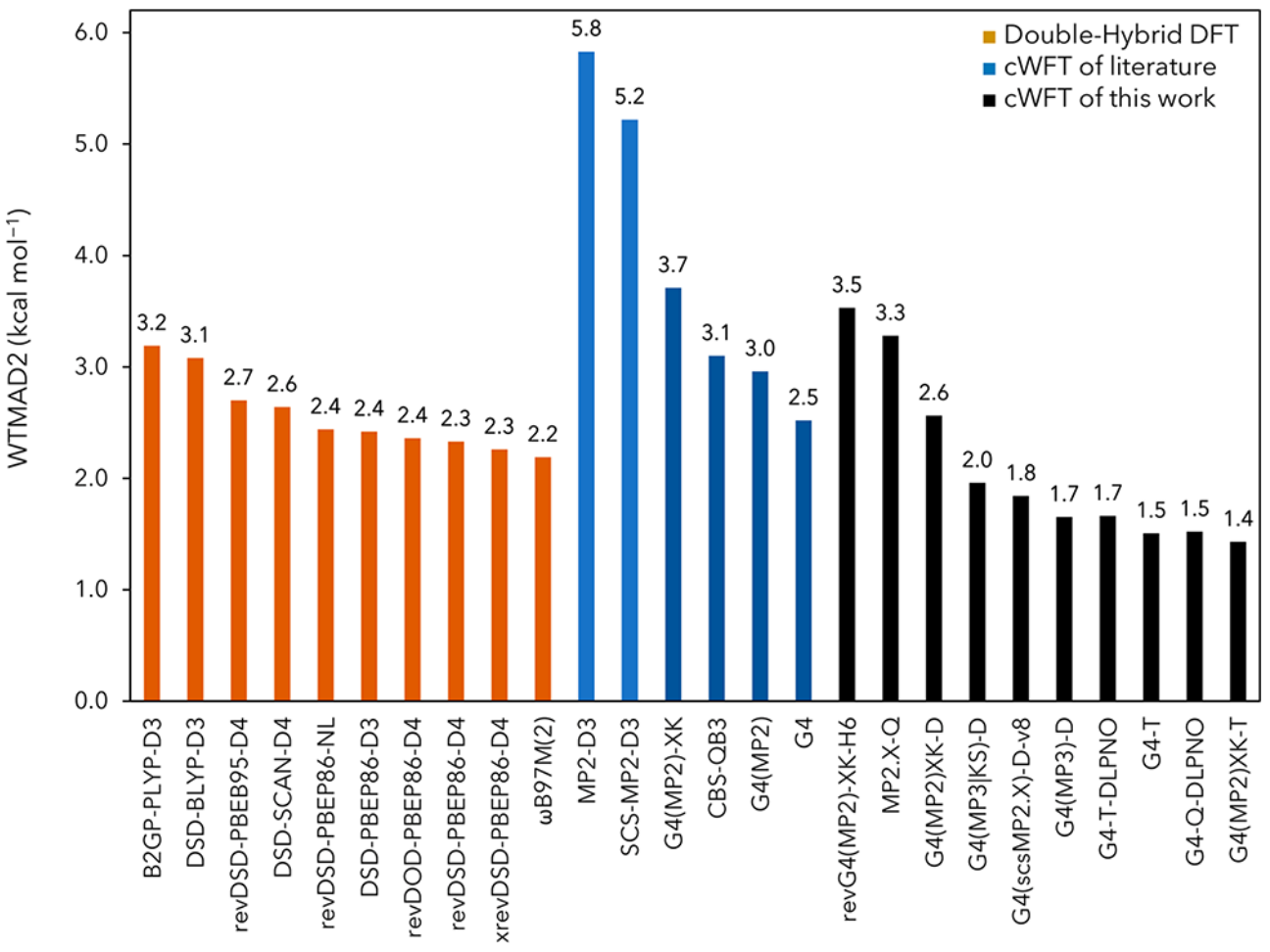

**Figure 2.** Weighted mean absolute deviation (WTMAD2 in kcal $mol^{-1}$) of selected composite wave function and double-hybrid DFT methods over the GMTKN55 database.

The best approximation has 13 empirical parameters. Can we reduce this number? Setting $A = A'$ comes at a minimal cost of just 0.02 kcal/mol. Instead eliminating the dispersion correction while retaining the HLC has a slightly greater cost, 0.07 kcal/mol, but then also setting $A = A'$ (leaving 11 parameters) barely affects statistics. (Conveniently, setting $A = A'$ eliminates a problem for radical reactions first identified by Chan, Coote, and Radom.[95]) However, deleting the HLC entirely and leaving just seven parameters yields a result of the same quality, 1.42 kcal/mol, which increases marginally to 1.43 kcal/mol if we remove the dispersion correction. Inspection of the WTMAD2 component breakdown reveals that dropping the HLC slightly degrades the thermochemistry and large-molecule reaction components but (as expected) leaves the intermolecular and conformer sets unaffected and actually slightly improves the barrier heights.

Extrapolating E2/{T,Q} instead [G4-T-v1 and G4-T-v2] gives WTMAD2 = 1.48 and 1.51 kcal/mol, respectively, with and without the dispersion correction: the latter especially could be regarded as a def2-based version of ccCA, albeit with the addition of three empirical parameters.

These latter statistics can be marginally lowered (about 0.02−0.03 kcal/mol) through separate extrapolation of $E_{2\alpha\beta}$ and $E_{2ss} \equiv E_{2\alpha\alpha} + E_{2\beta\beta}$, at the expense of introducing one additional parameter.

**IV.B. Using Three Basis Set Tiers.** Obviously, CCSD(T)/def2-TZVP is computationally quite costly for larger systems, and in fact proved out of reach for the very largest species in GMTKN55. Hence, we considered a three-tier approach akin to, but different from, G4 theory. The (scaled, see below) MP3 component was evaluated using the def2-TZVPP basis set, while for the CCSD(T)-MP3 difference, we fell back to the smallest basis set used in G4(MP2)-XK, namely, def2-SVSP, which is just def2-SVP but omitting the p polarization functions on hydrogen. For MP2, we used the same basis sets as in the previous paragraph.

Our best result here, WTMAD2 = 1.63 kcal/mol (G4-(MP3)-D-H6-v1), at first sight, represents a degradation of 0.3 kcal/mol over our best result in (III.A)—however, the comparison is not entirely fair as the 1.63 result is weighted over almost 200 additional (larger-molecule) reactions. If we re-evaluate for the same 1257 systems as above, then we find

WTMAD2 = 1.55, which represents about a 0.2 kcal/mol degradation for a vast reduction in CPU time.

By way of comparison, standard G4 theory—which employs a similar three-tier basis set scheme—achieves 2.52 kcal/mol on 1273 reactions. (It should be noted that this required tweaking the HLC for the rare gas atoms, as detailed at the end of the Methods section.) We note that the 2.52 kcal/mol number is in fact *inferior* to several of the best double-hybrid density functionals, notably[46] $\omega$B97M(2) of the Head–Gordon group (2.18 kcal/mol) and revDSD-PBEP86-D4 (2.33 kcal/mol) and xrevDSD-PBEP86-D4 (2.23 kcal/mol) of the Weizmann group. In contrast, our best approach here has comfortably lower WTMAD2 than all of these.

CBS-QB3, which is likewise a three-tiered method, but with MP4(SDQ) as the middle level, reaches WTMAD2 = 3.10 kcal/mol. This figure would be considerably lower were it not for the poor performance of the intermolecular interactions subsets (which at 1.55 kcal/mol accounts for half the total WTMAD2); detailed inspection reveals this to be due to basis set superposition error. Numerous DH-DFT methods are in the 2.3–2.7 kcal/mol WTMAD2 range, and they show significant overall improvement over CBS-QB3; i.e., $\omega$B97M(2) and xrevDSD-PBEP86-D4 are at 0.49 and 0.47 kcal/mol for the intermolecular interactions, respectively. At this point, we can correlate our CBS-QB3 findings with the conclusions of the detailed theoretical study of Karton and Goerigk on pericyclic reactions,[96] where dispersion was shown to play an important role (stabilizing the transition states). These authors compared the performance of WFT, cWFT, and DH-DFT methods with W1-F12 reference energies and found CBS-QB3 to be inferior (RMSD = 2.6 kcal/mol) to the DH-DFT methods DSD-BLYP-D3, B2GP-PLYP-D3, and B2K-PLYP-D3 (RMSD = 1.4 kcal/mol for all three).

Our best value requires 12 empirical parameters. Can we winnow this down at an acceptable cost in accuracy? Setting $A = A'$ costs a negligible 0.01 kcal/mol; excising the dispersion correction costs a still smallish 0.04 kcal/mol—but eliminating the HLC, thus cutting our number of parameters in half, actually achieves a slightly lower WTMAD2 at 1.65 kcal/mol. (This has the additional benefit of eliminating discontinuities in HLC as a bond is broken.) Once again, that small cost is confined to thermochemistry and large-molecule reactions.

Further reduction from six to five parameters can be achieved by eliminating the dispersion term, but now this comes at a cost of 0.11 kcal/mol. (Note the negative coefficient of −0.19 for the dispersion term, which appears to indicate it compensates for overbinding due to BSSE in the noncovalent interaction subsets. For detailed discussions of that issue, see refs 97 and 98.)

Using E2/{T,Q} extrapolation instead of scaling does come at a small cost in accuracy; however, the dispersion correction can then be eliminated at the expense of just 0.02 kcal/mol, leaving us with 1.84 kcal/mol (G4(scsMP2.X)-D-v8) for four parameters.

One additional refinement, at zero computational cost, we can consider would be to carry out separate extrapolations for $E_{2\alpha\beta}$ and $E_{2ss}$. Thus, introducing an additional parameter turns out to yield only negligible benefits across the board, and we have hence abandoned this.

The conventional MP3/def2-TZVPP turned out to be something of a bottleneck for the largest systems. This could obviously be mitigated with an RI-MP3 algorithm (e.g., such as reported in refs 99 and 100), but none was available to us. This prompted the question at what price the middle tier could be eliminated.

**IV.C. Eliminating the Middle Tier.** Our best result then (G4(MP2)-XK-D-H6-v1) is 2.52 kcal/mol, which represents a substantial degradation of over 1.63 kcal/mol above. A component breakdown (Table 1) reveals that the said degradation happens across the board for all subsets.

On the one hand, that implies the middle MP3-like tier (incidentally, inspired by the MP2.X method[101]) is clearly beneficial; on the other hand, that means that without it we are, in fact, at a slight disadvantage compared to the best double hybrids.

The costs of setting $A = A'$ in the HLC, and of eliminating the HLC outright, follow the same trends as in the previous section. This leaves us with 2.56 kcal/mol for seven parameters; additionally, dropping the dispersion correction brings us up to 2.73 kcal/mol.

The G4(MP2)-XK paper applies separate scaling coefficients to the overall CCSD correlation energy and the same-spin and opposite-spin CCSD components. (The third, hidden, component is the singles contribution for open-shell cases.) Does this breakdown have material advantages over simply scaling the CCSD-MP2 difference globally? We find [G4(MP2)-XK-D-H5-v1 vs G4(MP2)-D-H5-v1] that applying the latter constraint (thus eliminating two empirical parameters) adds just 0.06 kcal/mol to WTMAD2. Without dispersion [G4(MP2)-XK-D-H5-v2 vs G4(MP2)-D-H5-v2] this increases to 0.07 kcal/mol. In the absence of HLC, this increases further to 0.12 kcal/mol [G4(MP2)-XK-D-v1 vs G4(MP2)-D-v1], and when the dispersion correction is removed [G4(MP2)-XK-D-v2 vs G4(MP2)-D-v2] this shoots up to 0.3 kcal/mol. Hence, the advantage of scaling the CCSD spin components separately appears to lie primarily in noncovalent interactions. Setting $c_{\text{CCSD-MP2}} = c_{(\text{T})}$, i.e., treating all post-MP2 terms as a single correction, incurs a comparatively low cost. Somewhat surprisingly, perhaps, when eliminating (T) altogether (if dispersion is left in), WTMAD2 still stays below 3 kcal/mol. For perspective, however, we are now approaching the performance territory of the $\omega$B97M-V range-separated hybrid meta-GGA[91] (WTMAD2 = 3.29 kcal/mol),[46] the best rung-4 functional!

To put this in context, let us consider the original G4(MP2) and G4(MP2)-XK methods. For a smaller sample of 1208 molecules where we have G4(MP2) results, the WTMAD2 of G4(MP2) was equal to 2.96 compared to our best method, 2.31 kcal/mol.

(A short note is in order about some of the DFT functionals. WTMAD2 statistics for $\omega$B97X-V and $\omega$B97M-V were first published by Najibi and Goerigk,[102] and for B3LYP-D3 and M06-2X-D3(0) in the original GMTKN55 paper. These authors use the def2-QZVP basis set, augmented with diffuse s and p functions for the anionic subsets G21EA, AHB21, and IL16, while we, in the present work and in refs 45 and 46, employ def2-QZVPP as our baseline and def2-QZVPPD for the three aforementioned subsets plus RG18 and WATER27. While our WTMAD2 values (taken from ref 45 by way of ref 46) for $\omega$B97X-V and B3LYP-D3 are quite similar to those reported earlier in ref 102, our WTMAD2s for $\omega$B97M-V and M06-2X-D3(0) are lower by 0.24 and 0.40 kcal/mol, respectively. Part of the difference could be traced to our larger basis sets, and the remainder is likely due to the superfine integration grids used here: meta-GGAs are well-known to display strong grid sensitivity,[103] even though mitigation measures had been taken in the development of wB97M-V. For MP2-D3, the larger basis sets used here impart WTMAD2 improvements for the anionic species, but also for ACONF and RG18, which have large weights in the WTMAD2 formula.)

**IV.D. Reducing the MP2 Basis Set to TZ.** If we do so, we effectively obtain a reparametrized version (rev-G4MP2XK-H6-v1) of G4(MP2)-XK.[53] The refitted WTMAD2, 3.53 kcal/mol, is not much lower than the 3.71 kcal/mol obtained with the original G4(MP2)-XK parameters. This relatively small reduction suggests that the training set used in ref 53 was fairly representative. Both the original and refitted WTMAD2 values are, in fact, inferior to ωB97M-V and represent a marked deterioration over G4(MP2)-XK-T-H6-v1 in the previous section; a component breakdown of WTMAD2 (Table 1) reveals that this happens across the board. Considering the fairly low cost of an RI-MP2/def2-QZVPPD calculation if an RI-MP2 capable code is available (be it MOLPRO, Q-Chem, PSI4, ORCA, ...), there is really no valid reason not to "walk that additional tenth of a mile".

If we include the dispersion correction, WTMAD2 drops to 3.28 kcal/mol (note the sizable coefficient $c_{\text{Disp}} = -0.43$). Without D3(BJ), eliminating HLC comes at a price of about 0.5 kcal/mol, but with D3(BJ), no such large price need be paid: WTMAD2 = 3.35 kcal/mol, with $c_{\text{Disp}} = -0.81$. As above, the physical meaning of the large negative coefficient is clearly to compensate for basis set superposition error, particularly in the noncovalent interaction subsets.

**IV.E. Using the E2 Correlation Energy from a Double-Hybrid Calculation.** The good performance of double hybrids, inspired by GLPT2 (second-order Görling−Levy perturbation theory),[104] elicits the question: could we improve performance by using KS orbitals in the MP2 for the large-basis steps?

We have investigated this possibility both for the two-tier scenario from subsection IV.C and for the three-tier scenario from subsection IV.B. The E2 energy components were taken from a ωrevDSD-PBEP86-D3BJ calculation with $x = 0.72$, $\omega = 0.16$.

In the two-tier scenario, we see no noticeable improvement, but an intriguing phenomenon can be observed in the triple excitations coefficient $c_T$, which takes on relative small *negative* values. In the three-tier scenario (G4(MP3|KS)-D-H5-v1), we can get down to WTMAD = 1.89 kcal/mol—but $c_T \approx 0$, and in addition $c_{\text{CCSD−MP3}} \approx 0$. This offers the tantalizing possibility of eliminating the costly CCSD(T) calculation step entirely (i.e., $c_{\text{CCSD-MP3}} = c_T = 0$): doing so [G4(MP3|KS)-D-H5-v5 and G4(MP3|KS)-D-H5-v6] yields WTMAD = 1.89 with or without $c_{\text{Disp}} = 0$. This offers a performance superior to double hybrids but requiring no steps more expensive than MP3/def2-TZVPP and with only three parameters (plus the six of the HLC). Can we cut out the HLC as well? Doing so [G4(MP3|KS)-D-v5 and G4(MP3|KS)-D-v6] only slightly increases WTMAD2 to 1.93 kcal/mol (with dispersion) and 1.96 kcal/mol (without dispersion): these last two options only have four and three parameters total, respectively, i.e., fewer than the double hybrids.

We note that code limitations precluded carrying out MP3 calculations in a basis of KS orbitals; hence, the MP3 term could only be considered in a basis of HF orbitals. From another perspective, that of fifth-rung density functional methods, this points toward a path for further improving on ωB97M(2) and revDSD-PBEP86-D4 at a relatively modest computational cost. Past attempts involving post-MP2 methods by Chan, Goerigk, and Radom,[105] using correlation ranging from MP3 to CCSD(T), showed no improvement: however, these authors for obvious cost reasons used the G2/97 small-molecule thermochemistry data set of 148 small molecules (about one-tenth the size of the present training set) and also restricted themselves to TZ quality basis sets. In view of our findings about the importance of a QZ quality basis set for the MP2-like component, it is quite possible that any improvements from post-MP2 methods in ref 105 were masked by residual basis set incompleteness error in the KS-MP2 contribution. The proverb about a chain being no stronger than its weakest link comes to mind.

The benefits of using KS orbitals in the MP2 for the large-basis steps can be illustrated from a different angle: if we use instead (spin-component-scaled) MP2, MP3, and a dispersion correction and optimize the four adjustable parameters (MP2.X-Q), we obtain WTMAD2 = 3.28 kcal/mol, which stays the same if we cut out the dispersion correction. Comparison with the MP2|KS cognates of these two methods (1.93 and 1.96 kcal/mol, see above) reveals that substituting MP2|KS for MP2 imparts a benefit of about 1.2−1.3 kcal/mol. A component breakdown shows that the difference lies in the small-molecule thermochemistry, barrier heights, and large-molecule reaction energies, while the two noncovalent interaction subsets taken together stay approximately unchanged in quality.

For this "empirical SCS-MP2x" approach, basis set extrapolation does offer a slight improvement (on the order of 0.05 kcal/mol).

**IV.F. Performance of the DLPNO-CCSD(T)-Based Composite Methods.** As the O($n^7$) CPU time scaling with a system size of canonical CCSD(T) will present an insurmountable obstacle for application to larger systems, we considered its replacement by the DLPNO-CCSD(T) approach, which is linear-scaling with system size. Specifically, for two-tier approaches, we replaced the post-MP2 correction, $E$[CCSD(T)] − $E$[MP2], by $E$[DLPNO-CCSD(T)] − $E$[DLPNO-MP2] and left the remaining terms unchanged. The performance of such methods, in which we combine an E2/{T,Q} extrapolation with post-MP2 corrections for different-sized basis sets, is illustrated in Figure 3. Using the largest basis set def2-QZVPP in the post-MP2 correction, our best overall WTMAD2 is 1.50 kcal/mol for G4-Q-DLPNO-H6-v1 with 10 parameters. Setting $c_{\text{Disp}} = 0$ for G4-Q-DLPNO-H6-v2 leaves WTMAD2 unchanged. When $A = A'$ for HLC, that yields 1.51 kcal/mol with or without dispersion [G4-Q-DLPNO-H5-v1 and G4-Q-DLPNO-H5-v2]. Eliminating the HLC has a negligible effect: WTMAD2 = 1.52 kcal/mol for G4-Q-DLPNO-v1 and G4-Q-DLPNO-v2. The second variant, G4-Q-DLPNO-v2, has only three parameters, and this minimally empirical cWFT will be referred to as G4-Q-DLPNO.

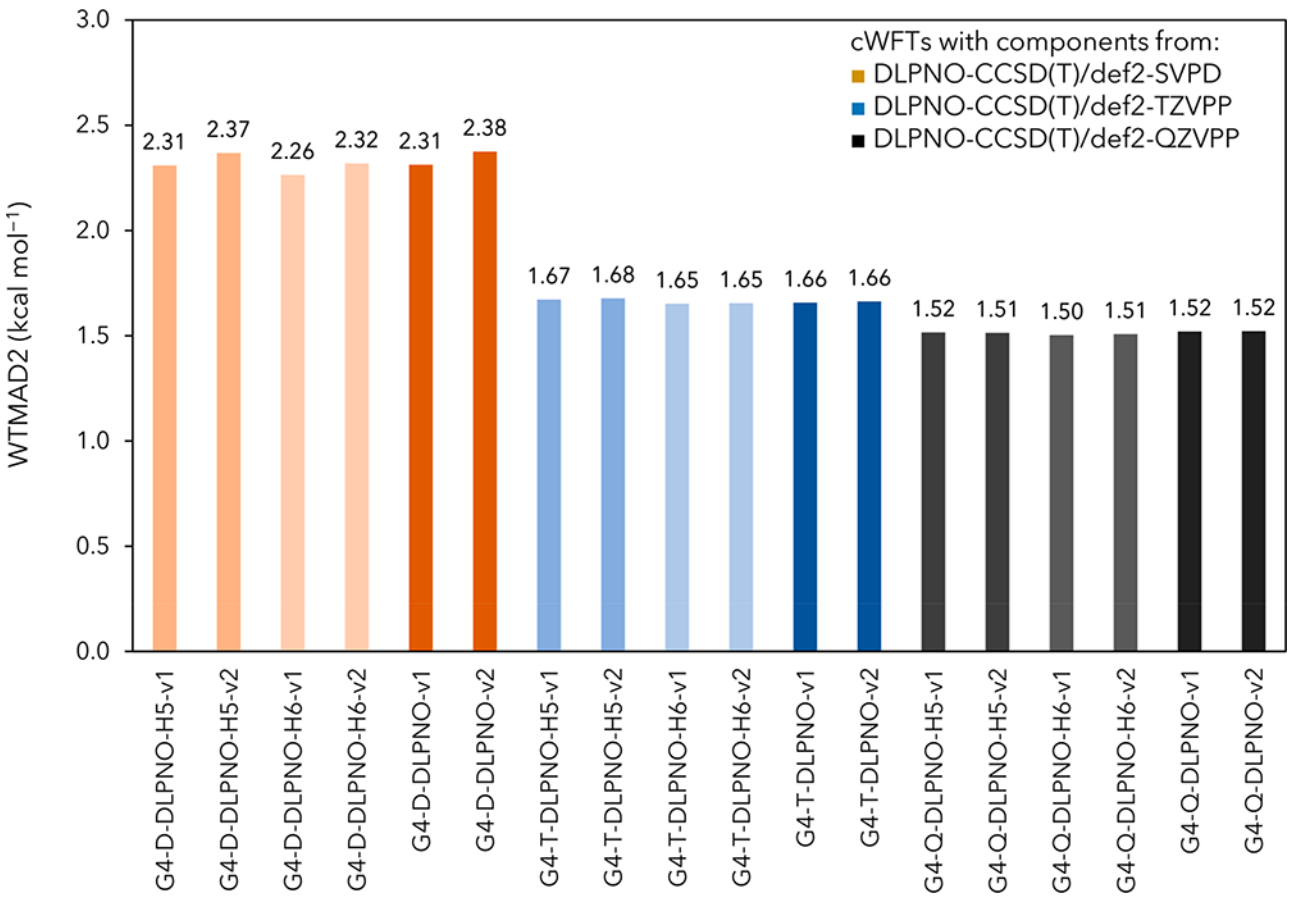

**Figure 3.** Weighted mean absolute deviation (WTMAD2 in kcal mol$^{-1}$) of DLPNO-CCSD(T)-based composite methods over the GMTKN55 database. Darkest colors represent the non-HLC variants.

What about using the smaller basis sets def2-TZVPP or even def2-SVPD in the DLPNO-CCSD(T) step instead? For the larger def2-TZVPP, the dispersion similarly has a negligible contribution to WTMAD2 of all G4-T-DLPNO variants. The lowest WTMAD2 is 1.65 for G4-T-DLPNO-H6-v1 with 10 parameters and does not change when constraining $c_{\text{Disp}} = 0$. Setting $A = A'$ slightly raises WTMAD2 by 0.03 with, and 0.04 kcal/mol without, dispersion [G4-T-DLPNO-H5-v1 and G4-T-DLPNO-H5-v2, respectively]. Removing HLC yields 1.67 and 1.68 kcal/mol when retaining or discarding dispersion, respectively.

The smallest basis set def2-SVPD renders the dispersion beneficial for the G4-D-DLPNO variants. The lowest WTMAD2 is 2.26 kcal/mol for G4-D-DLPNO-H6-v1 and 10 fitted parameters. Removal of dispersion increases WTMAD2 by 0.06 kcal/mol, while setting $A = A'$ raises it by 0.05 and 0.10 kcal/mol with and without dispersion [G4-D-DLPNO-H5-v1 vs G4-D-DLPNO-H5-v2]. Eliminating the HLC increases WTMAD2 to 2.31 kcal/mol for G4-Q-DLPNO-v1, while further eliminating the dispersion leads to WTMAD2 = 2.38 kcal/mol for G4-D-DLPNO-v2 with only three parameters. This is better by 1.33 kcal/mol than the original G4(MP2)-XK at a fraction of its cost, and with nine fewer parameters.

Recently, Neese and co-workers[106] found for the GMTKN55 data set and the aug-cc-pVDZ basis set that DLPNO-CCSD($T_1$) with TightPNO has a WTMAD2 from the canonical results of 0.87 kcal/mol (1.58 kcal/mol for NormalPNO). Intriguingly, the present results show a much smaller difference between corresponding canonical and DLPNO approaches, implying a degree of error compensation between the MP2 and post-MP2 components. Moreover, replacing DLPNO-CCSD(T) by DLPNO-CCSD($T_1$) yields a marginal improvement of statistics for the G4-DLPNO-T1 type methods in this framework (see the Supporting Information).

Gas-phase thermochemistry of small organic molecules has previously been studied employing DLPNO-CCSD(T) electronic energies by Paulechka and Kazakov,[107] who compared experimental gas-phase formation enthalpies with those at DLPNO-CCSD(T)/def2-QZVP//RI-MP2/def2-QZVP. They considered four variants, the most costly scheme being single-point energies at DLPNO-CCSD(T)/def2-QZVP and a geometry optimization at RI-MP2/def2-QZVP. Their data set consisted of 45 small organic molecules, ranging in size from benzene and pyridine to biphenyl. That implementation was available only for C, H, O, and N containing closed-shell molecules. Mielczarek et al.[108] extended this work to 164 molecules composed of the elements H, C, N, O, F, S, Cl, and Br.

**IV.G. Final Selected Methods.** It is clear from the discussion above that the WTMAD2 improvement from including the HLC is not commensurate with having to include an extra six parameters. This winnows our choices to the non-HLC ones.

We then find that the benefit of the dispersion corrections is negligible for the QZ/TZ two-tier approaches but rather less negligible for the QZ/TZ/DZ three-tier approaches and the Q/D two-tier. This leaves us with the following hierarchy of three methods; the variant label is omitted for the final selected methods, and the original method's name is in brackets:

GOOD: [G4(MP2)-XK-D-v1] **G4(MP2)-XK-D WTMAD2 = 2.56 kcal/mol**, with seven fitted parameters, and with CCSD(T)/def2-SVSP as the computationally costliest step;

BETTER: [G4(MP3)-D-v1] **G4(MP3)-D WTMAD2 = 1.65 kcal/mol** with six fitted parameters, and with MP3/def2-TZVPP and CCSD(T)/def2-SVSP as the costliest steps;

BEST: [G4(MP2)-XK-T-v2] **G4(MP2)-XK-T WTMAD2 = 1.43 kcal/mol** likewise with six fitted parameters, and with CCSD(T)/def2-TZVP as the costliest step. (The "T" refers to the basis set for CCSD(T)/def2-TZVP used there. The basis set extrapolation steps are only a subsidiary element of the computational cost if an RI-MP2 approximation is used.)

If one seeks an alternative to "best" with fewer fitted parameters, [G4-T-v2] **G4-T** with **WTMAD2 = 1.51 kcal/mol** fits the bill.

If KS orbitals in the MP2 are used, then [G4(MP3|KS)-D-v6] **G4(MP3|KS)-D** with **WTMAD2 = 1.96 kcal/mol**, with no step costlier than MP3/def2-TZVPP.

If one considers DLPNO-CCSD(T) as a gentler-scaling alternative to canonical CCSD(T), then G4-T-DLPNO becomes an attractive option, with **WTMAD2 = 1.66 kcal/mol**, no step costlier than DLPNO-CCSD(T)/def2-TZVPP, and just three fitted parameters. Somewhat superior accuracy can be obtained (**WTMAD2 = 1.52 kcal/mol**) at the **G4-Q-DLPNO** level, where DLPNO-CCSD(T)/def2-QZVPP is the costliest step.

In Table 3, we present the final selected cWFT methods. They combine the highest accuracy for the molecular energies across the GMTKN55 database with a minimal number of empirical parameters. The energy expressions that we employed eliminate any redundant parameters (HLC) and render the listed methods transferable to various chemical systems, other than those of the 2459 molecules of GMTKN55. Extensive benchmark sets and the survival of the fittest approach, where the proper energy expressions eliminate empirical parameters, improved the DFT performance significantly within chemical accuracy for energetics. Representative cases include ωB97M(2) and revDSD-PBEP86-D4 that have been successfully employed to expanded porphyrins[109] or transition metal reaction barrier heights of the MOBH35 database.[110,111]

Example input and output files for the popular Gaussian 16 electronic structure system, as well as postprocessing scripts, have been given in the Supporting Information. The geometry optimization and frequencies steps (and scale factors) are the same there as in the original G4(MP2)XK specification.[53]

When perusing the CPU and wall clock times in these outputs, it should be kept in mind that this code has no RI-MP2 implementation, and hence the largest basis set MP2 step will weigh much heavier on the total than it would with another code like ORCA or Q-CHEM.

Further performance assessment of the presently proposed cWFT methods is especially desirable for transition metal (and heavy p-block) energetics and barrier heights, particularly in order to learn their limitations. Toward that aim, we shall consider the following data sets in future work: TMC151 of transition metals;[112] CUAGAU of small copper, silver, and gold compounds;[113] the MOBH35 (metal−organic barrier height) data set;[110,111] and for the main group, the heavy p-block subset from ref 53 and the smaller MG8 of noncovalent interactions, isomerization reactions, thermochemistry, and barrier heights.[114]

## V. CONCLUSIONS

We have attempted to develop a hierarchy of composite WFT computational thermochemistry protocols based on Weigend−Ahlrichs basis sets but parametrized against the very large and chemically diverse GMTKN55 benchmark.

The original G4(MP2)-XK approach of Chan, Karton, and Raghavachari could be considered the lowest tier of our approach, using no larger basis set than def2-TZVPP for the MP2 correlation and def2-SVSP for the post-MP2 corrections. For GMTKN55, with the original parameters (but including a fix for rare gas atoms), G4(MP2)-XK is intermediate in performance between the ωB97X-V and ωB97M-V combinatorially optimized range-separated hybrid functionals. Refitting lowers WTMAD2 somewhat, but not spectacularly. Adding an empirical dispersion correction brings G4(MP2)-XK-D in the WTMAD2 = 3.3 kcal/mol range (equal to ωB97M-V, the best-performing rung-4 functional) and additionally allows us to eliminate the "high-level correction".

If an RI-MP2 code is available, expanding the MP2 basis set to def2-QZVPPD is relatively inexpensive in both CPU time and storage requirements. The resulting G4(MP2)-XK-D method performs in the same range as revDSD double hybrids but is still inferior to the best among them like revDSD-PBEP86-D4, as well as to the combinatorially optimized ωB97M(2) range-separated double hybrid. Especially if a dispersion correction is included, the HLC turns out to be superfluous, bringing the total number of empirical parameters down to just six.

A significant improvement is achieved with the three-tier G4(MP3)-D method, which entails RI-MP2/def2-QZVPPD, MP3/def2-TZVPP, and CCSD(T)/def2-SVSP calculations. We can then achieve WTMAD2 = 1.65 kcal/mol, markedly superior to even the best double hybrids.

A further enhancement is possible as a two-tier approach with RI-MP2/def2-QZVPPD and CCSD(T)/def2-TZVP, but at WTMAD2 = 1.43 kcal/mol, the substantial premium in computational cost is hard to justify by a performance benefit of just 0.2 kcal/mol over G4(MP3)-D. Still, further improvements probably will require going to quintuple-zeta basis sets for the MP2 step, including inner-shell correlation, considering post-CCSD(T) correlation effects, or a combination thereof.

While G4(MP2)-XK-D and G4(MP3)-D are comparable in computational cost to G4(MP2) and G4, respectively, their WTMAD2 metrics over GMTKN55 are markedly superior. A component breakdown of WTMAD2 reveals that the advantage of G4(MP2)-XK-D and G4(MP3)-D resides primarily in the large-molecule, intramolecular NCI (non-covalent interactions), and intermolecular NCI subsets: we note that the HLC as defined for G4(MP2) and G4 cancels exactly between the reactant and product sides of these reactions (as for any reactions that only involve closed-shell molecules), and hence it is unable to correct for basis set incompleteness.

The combination of (MP2|KS)/def2-QZVPPD correlation from a double-hybrid calculation with scaled MP3/def2-TZVPP correlation energy yields very promising results, WTMAD2 = 1.96 kcal/mol, at a comparatively low computational cost. This points toward an avenue for further improving the performance of empirical double-hybrid density functionals.

The DLPNO-CCSD(T)-based variants of the two-tier family represent the lowest-cost approach due to their minimal requirements compared to the canonical CCSD(T)-based cWFT. (By way of illustration: the most time-consuming step of G4-T-DLPNO on a melatonin conformer took about 2 h on eight cores of a Skylake machine, compared to about 2 days for the canonical G4-T. This gap will only widen for larger molecules.) G4-D-DLPNO was found to have WTMAD = 2.38 kcal/mol, i.e., 1.33 kcal/mol lower that the original G4(MP2)-XK at a fraction of its cost and with nine fewer parameters. The G4-Q-DLPNO and G4-T-DLPNO had the lowest WTMAD2 = 1.52 and 1.65 kcal/mol for the GMTKN55 database, respectively. Out of the different DLPNO-based approaches, G4-T-DLPNO appears to offer the best price−performance ratio.

Overall, we can conclude that composite WFT methods still offer advantages over the best double hybrids—but that the gap has narrowed substantially.

Finally, we would like to point out that G4-T, G4-Q-DLPNO, G4-T-DLPNO, G4(MP2)-XK-T, G4(MP2)-XK-D, G4(MP3)-D, and G4(MP3|KS)-D are available for all s, p, and d block elements of the Periodic Table. Sample inputs and postprocessing scripts are given in the Supporting Information.

## ■ ASSOCIATED CONTENT

### Supporting Information

The Supporting Information is available free of charge at https://pubs.acs.org/doi/10.1021/acs.jctc.0c00189.

Gaussian 16 files (including required basis sets) and postprocessing scripts for the G4-T, G4(MP2)-XK-T, G4(MP3|KS)-D, G4(MP2)-XK-T, and G4(MP3)-D methods, accompanied by an instructions document; detailed equations for all composite methods developed here; the full table of parameters and statistics of all methods considered in Microsoft Excel format; and a report with a detailed breakdown of all 55 individual subsets of the performance statistics of the methods proposed in this paper (ZIP)

## ■ AUTHOR INFORMATION

### Corresponding Author

**Jan M. L. Martin** − *Department of Organic Chemistry, Weizmann Institute of Science, 7610001 Reḥovot, Israel;* orcid.org/0000-0002-0005-5074; Email: gershom@weizmann.ac.il; Fax: +972 8 9343029

### Author

**Emmanouil Semidalas** − *Department of Organic Chemistry, Weizmann Institute of Science, 7610001 Reḥovot, Israel;* orcid.org/0000-0002-4464-4057

Complete contact information is available at: https://pubs.acs.org/10.1021/acs.jctc.0c00189

### Author Contributions

The manuscript was written through contributions of all authors. All authors have given approval to the final version of the manuscript.

### Funding

Research was supported by the Israel Science Foundation (grant 1358/15) and by the Yeda-Sela-SABRA program (Weizmann Institute of Science). The work of E.S. on this scientific paper was supported by the Onassis Foundation—Scholarship ID: F ZP 052-1/2019-2020.

### Notes

The authors declare no competing financial interest.

## ■ ACKNOWLEDGMENTS

J.M.L.M. acknowledges Dr. Mark Vilensky (scientific computing manager of ChemFarm) and Mr. Jacov Wallerstein (CEO and CTO of Access Technologies) for development of the bscratch shared high-bandwidth scratch storage server, without which some of the largest conventional calculations would have been impossible. Peer reviewers are thanked for helpful suggestions. We also thank Mr. Golokesh Santra and Dr. Mark A. Iron for helpful discussions.

## ■ REFERENCES